\documentclass[aps,pra,twocolumn,superscriptaddress,amsmath,amssymb]{revtex4-2}

\usepackage{graphicx}
\graphicspath{{Figure/}}
\usepackage{siunitx}
\usepackage{upgreek}
\UseRawInputEncoding

\newcommand{\sect}[1]{\par\textit{#1}.---\ignorespaces}
\newcommand{\void}[1]{}

\begin{document}

\title{Microwave-resonator-detected excited-state spectroscopy of a double quantum dot}


\author{Ming-Bo Chen}
\author{Shun-Li Jiang}
\author{ Ning Wang}
\author{Bao-Chuan Wang}
\author{Ting Lin}
\author{Si-Si Gu}
\author{Hai-Ou Li}
\affiliation{Key Lab of Quantum Information, CAS, University of Science and Technology of China, Hefei, China}
\affiliation{CAS Center for Excellence in Quantum Information and Quantum Physics, University of Science and Technology of China, Hefei, Anhui 230026, China}

\author{Gang Cao}
 \email{gcao@ustc.edu.cn}
\affiliation{Key Lab of Quantum Information, CAS, University of Science and Technology of China, Hefei, China}
\affiliation{CAS Center for Excellence in Quantum Information and Quantum Physics, University of Science and Technology of China, Hefei, Anhui 230026, China}

\author{Guo-Ping Guo}
\email{gpguo@ustc.edu.cn}
\affiliation{Key Lab of Quantum Information, CAS, University of Science and Technology of China, Hefei, China}
\affiliation{CAS Center for Excellence in Quantum Information and Quantum Physics, University of Science and Technology of China, Hefei, Anhui 230026, China}
\affiliation{Origin Quantum Computing Company Limited, Hefei, Anhui 230026, China}

\begin{abstract}
As an application in circuit quantum electrodynamics (cQED) coupled systems, superconducting resonators play an important role in high-sensitivity measurements in a superconducting-semiconductor hybrid architecture. 
Taking advantage of a high-impedance NbTiN resonator, we perform excited-state spectroscopy on a GaAs double quantum dot (DQD) by applying voltage pulses to one gate electrode. 
The pulse train modulates the DQD energy detuning and gives rise to charge state transitions at zero detuning. 
Benefiting from the outstanding sensitivity of the resonator, we distinguish different spin-state transitions in the energy spectrum according to the Pauli exclusion principle. 
Furthermore, we experimentally study how the interdot tunneling rate modifies the resonator response. The experimental results are consistent with the simulated spectra based on our model.
\end{abstract}

\date{\today}
\maketitle

Semiconductor quantum dots are promising candidates for quantum computation \cite{Kloeffel2013a,Zhang2019a}. 
Conventionally, qubits defined in quantum dots are studied by direct current transport measurements 
\cite{Tarucha1996a,Watzinger2018a,Hendrickx2020a}, 
but the readout fidelity is limited by the small tunneling current in the few-electron regime. 
External electrometers such as quantum point contacts and single electron transistors are also commonly used in quantum dot devices
\cite{Watson2018a,Mills2019a,Assad2020a,Broome2018a},
but the short-range Coulomb interaction will be an obstacle for the electrode layout when scaling up \cite {Vandersypen2017a,Veldhorst2017a,Jones2018a}. 
In addition, both readout methods inevitably introduce coupling to the environment that damps the quantum state coherence
\cite{Bermeister2014a,Martins2016a,Dial2013a}, 
or leads to back action that causes deleterious inelastic transitions between the quantum dots and the leads   \cite{Granger2012a,Cao2013a}. 

Alternatively, gate-based electrometers using on-chip superconducting resonators \cite{Zheng2019a,Scarlino2019a} 
and lumped element resonators \cite{Colless2013a,West2019a,Crippa2019a,Schaal2020a} 
have been developed with high readout fidelity. In these designs, only a single gate electrode is required for readout, significantly simplifying the gate structure. 
In particular, resonators have the advantage of performing quantum non-demolition readout 
\cite{Wallraff2005a} and can be utilized as mediators for long-distance coupling and entanglement between qubits, which provide a desirable route for scalable semiconductor quantum computing
 \cite{vanWoerkom2018a,Borjans2020a}.

In recent years, research on circuit quantum electrodynamics (cQED) hybrid devices with semiconductor quantum dots has been greatly promoted with the help of high-impedance resonators that improve the coupling strength 
\cite{Stockklauser2017a,Landig2018a,Mi2018a,Samkharadze2018a,Wang2020a}.   
With the application of high-impedance resonators, various measurements for characterizing qubits have been performed, such as detecting of stability diagram, spectroscopy of valley states \cite{Burkard2016a}
and distinguishing of two-electron spin states \cite{Landig2019a}. 

Here, we use a high-impedance NbTiN superconducting resonator to perform excited-state spectroscopy on a GaAs double quantum dot (DQD). 
The energy spectrum is dispersively measured by applying a train of 50\% duty-cycle square pulses to the DQD in the four-electron regime  \cite{Elzerman2004a,Johnson2005b,Shi2014a,Chen2017a}. 
According to the resonator response, we can probe the pulse-induced electron-state transitions \cite{Burkard2016a,Zheng2019a} 
and identify the dynamics of interdot transitions between different spin and orbital states. 
Utilizing the tunability of the semiconductor DQD,
we further investigate the DQD parameter dependence of the resonator response and numerically calculate the influence of the interdot tunneling rate. 
The combination of low-frequency pulsed-gate technology and advanced gate-based photonic probe method provides a powerful tool for detecting singlet-triplet qubits and characterizing multi-level quantum dot systems.

\begin{figure}[bhp]
\includegraphics[width=\columnwidth]{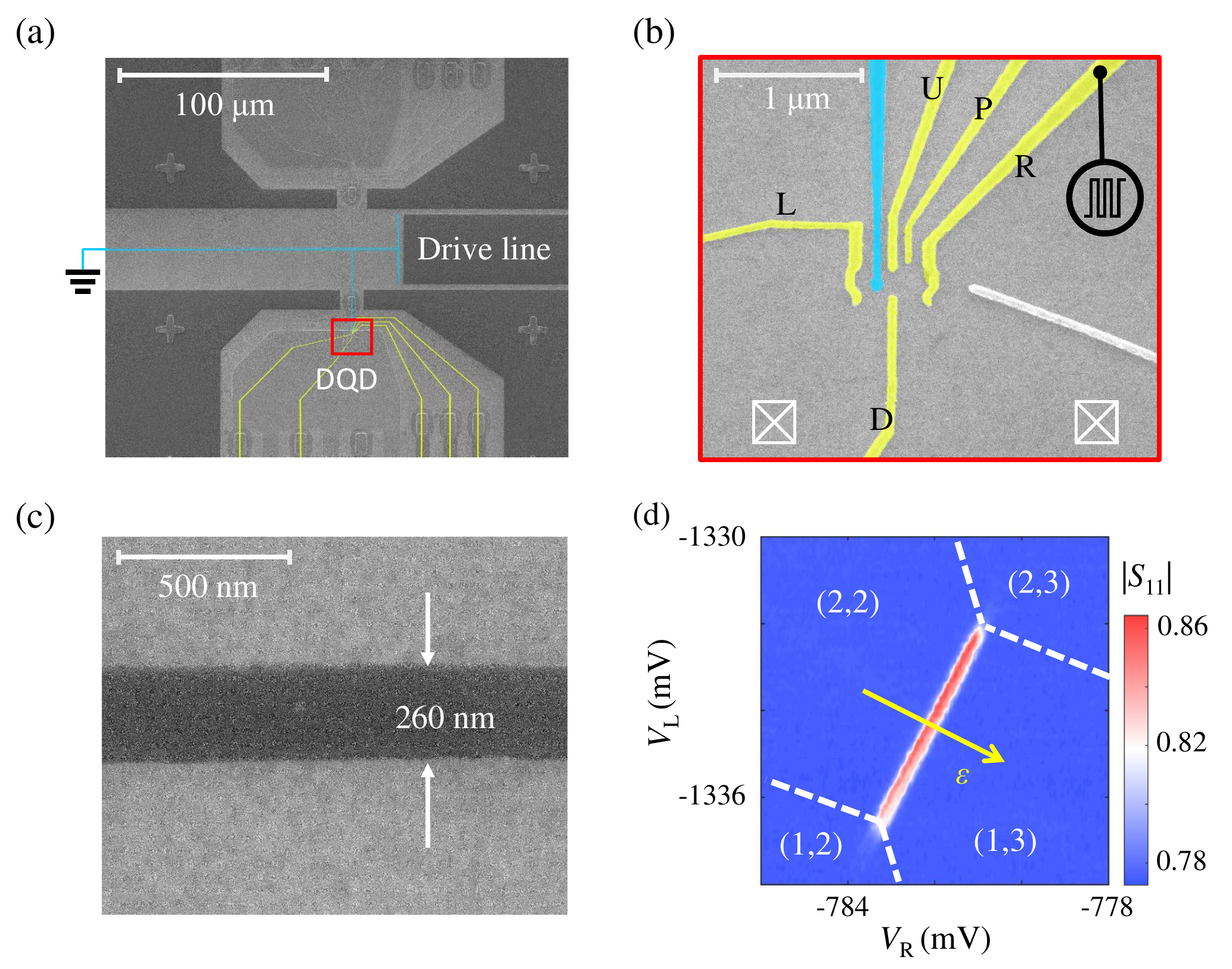} 
\caption{Hybrid device and stability diagram. 
(a) False-color scanning electron micrograph of the device, showing the resonator (azure) and gate electrodes (yellow) of the DQD. 
The resonator  capacitively couples to the drive line at one end  and grounded at the other end. 
The DQD region is indicated by a red rectangle.  
(b) DQD defined by metallic top gates. The left plunger gate highlighted in blue is connected to the NbTiN superconducting resonator. 
Pulses are applied to gate R to perform the excited-state spectroscopy. 
(c) NbTiN resonator with a narrow width of $\SI{260}{\nm}$. 
(d) Transition line between charge configurations (2,2) and (1,3) measured by the resonator. 
The detuning $\varepsilon$ indicated by an arrow can be controlled by the gate voltage $V_{\text R}$. 
Dashed lines indicate the electron tunneling between the DQD and the leads. }
\label{fig:setup}
\end{figure}

\sect{Setup}
As shown in Fig.~\ref{fig:setup}(a)--(c), the device is composed of a DQD and a quarter-wavelength superconducting resonator, fabricated on a GaAs/AlGaAs heterostructure. 
The two-dimensional electron gas is removed everywhere except for a small mesa region that hosts the DQD. 
The DQD is defined by applying DC voltages to top gate electrodes L, R, U and D. 
Gates L and R control the tunneling barriers between the dots and the leads. Gates U and D control the interdot tunneling rate $2t$. 
The plunger gate labeled P is zero biased and idle.
In our measurement, the source and drain are grounded. 

To couple the resonator photons to the qubit, the left plunger gate of the DQD is  galvanically connected to the voltage antinode of the superconducting resonator. 
The resonator is fabricated from 11-nm-thick NbTiN  with a narrow center conductor and remote ground planes using electron-beam lithography, deposited on a substrate with a 15-nm-thick Al\textsubscript{2}O\textsubscript{3} layer. 
The center conductor, with a width of $\SI{260}{\nm}$ and a length of $ \SI{300}{\um}$, is grounded at one end and capacitively coupled to a drive line at the other end. 
The choice of NbTiN material and extremely small cross section 
\cite{Samkharadze2016a} leads to a large kinetic inductance $L_{\text k} \approx \SI{320}{\pico\henry\per\um }$. 
Then the characteristic impedance $Z_{\text r} \approx \SI{2.5}{\kohm}$ is far beyond that of conventional $\SI{50}{\ohm}$ resonators \cite{Petersson2012a,Deng2015a}. 
This high-impedance resonator helps improve the coupling strength $g_0$ between the DQD and resonator which is proportional to $\sqrt{Z_{\text r}}$ \cite{Childress2004a}.

Our device is cooled in a dilution refrigerator with a base temperature of $\sim \SI{23}{\milli\kelvin}$. 
We measure the response of the resonator using a vector network analyzer with probe frequency $f_{\text p}$. 
The microwave photons are attenuated approximately $\SI{70}{\dB}$ before reaching the drive line, reflected by the resonator, and then amplified at $\SI{4}{\kelvin}$ and room temperature sequentially. 
Measuring the reflection signal $S_{11}$, we determine that the resonator frequency is $f_{\text r}=\omega_{\text r}/2\pi=\SI{6.04}{\GHz}$ with a photon decay rate of $\kappa/2\pi=\SI{35.3}{\MHz}$.

This cQED hybrid system can be described by the Jaynes-Cummings model \cite{WallsMilburn1995a} with the  Hamiltonian 
\begin{equation}
H=\hbar\Omega\sigma_\text{z}/2+\hbar \omega_{\text r} a^{\dag} a+\hbar g_{\text c} (a^{\dag} \sigma_{-}+a\sigma_{+} ),
\label{Hamiltonian}
\end{equation}
where $\hbar\Omega=\sqrt{\varepsilon^2+(2t)^2}$ is the energy difference between the lowest two charge states of the DQD, 
$\varepsilon$ is the energy detuning,  
$\sigma_\text{z}$ and $\sigma_{\pm}$ are the Pauli operators in the DQD eigenstate basis, $a^{\dag}$ and $a$ are the photon creation and annihilation operators, 
$g_{\text c}$ is the effective coupling strength given by $g_{\text c}=g_0\cdot 2t/\hbar\Omega$ \cite{Childress2004a},
with $g_0$ being the overall coupling rate, and $\hbar$ is the reduced Plank constant. 
The first (second) term represents the free Hamiltonian of the qubit (resonator). 
The third term describes the charge-cavity electric dipole coupling. 
Due to the coupling rate $g_{\text c}$, 
a change in the DQD charge state results in a frequency shift $\Delta f_{\text r}$ 
on the resonator, varying the amplitude and phase of the reflection signal \cite{Petersson2012a,Deng2015a}. 
Specifically, in the dispersive limit $g_{\text c}\ll \Delta$, where $\Delta =\Omega-\omega_{\text r}$ is the detuning between the DQD and resonator, the Hamiltonian can be simplified to 
\begin{equation}
H\approx \hbar\Omega\sigma_\text{z} /2+\hbar(\omega_{\text r}+\frac{g_{\text c}^2}{\Delta}\sigma_\text{z})a^{\dag} a,
\label{dispersive}
\end{equation}
and the frequency shift $\Delta f_{\text r}$ is proportional to 
$g_{\text c}^2/\Delta$.
Therefore, we can probe the tunneling dynamics of the DQD according to the resonator response.

The DQD is configured and characterized by the resonator reflection amplitude $|S_{11}|$ in response to the gate voltages. 
Figure~\ref{fig:setup}(d) shows the charge stability diagram near the (2,2)--(1,3) charge transition obtained by measuring $|S_{11}|$ as a function of $V_{\text L}$ and $V_{\text R}$ with probe frequency $f_{\text p}=f_{\text r}$, where $(m,n)$ denotes $m~(n)$ electrons in the left (right) quantum dot. 
The detuning $\varepsilon$, indicated by a yellow arrow in the diagram, is defined as the electrochemical potential difference between the two dots.
It can  be controlled by gate R with a voltage-to-energy conversion ratio of $\alpha/h\approx \SI{20.8}{\GHz/\mV}$ \cite{Kouwenhoven1994a}. 
With appropriate gate voltages, the electrochemical potentials of the two dots are aligned and $\varepsilon=0$,
leading to a coherent interdot transition that increases  $|S_{11}|$. 
Fitting $|S_{11}|$ with the input-output theory \cite{Viennot2015a,Mi2018a,Landig2019a}, we obtain the coupling strength $g_0/2\pi\approx\SI{104.5}{\MHz}$, the qubit decoherence rate $\gamma/2\pi\approx\SI{72}{\MHz}$ and the tunneling rate $2t/h\approx\SI{5.8}{\GHz}$, indicating that the DQD-resonator system reaches the strong coupling limit \cite{Stockklauser2017a,Landig2018a,Mi2018a,Samkharadze2018a,Wang2020a}.
This large coupling strength, benefiting from the high-impedance resonator, enables dispersive probing of the DQD state with high resolution.
Since the two core electrons in the right dot occupy the ground orbit and form an "isolated" shell, their contribution to the dynamics is ignored in our analysis.
The relevant physics is considered the same as that at the (2,0)--(1,1) transition \cite{Cao2016a,Yang2020a,Petit2020a}.
In the following, we use a two-electron configuration to describe the pulse-induced dynamics in this four-electron regime.  

\begin{figure*}[htb]
\includegraphics[width=17.8cm]{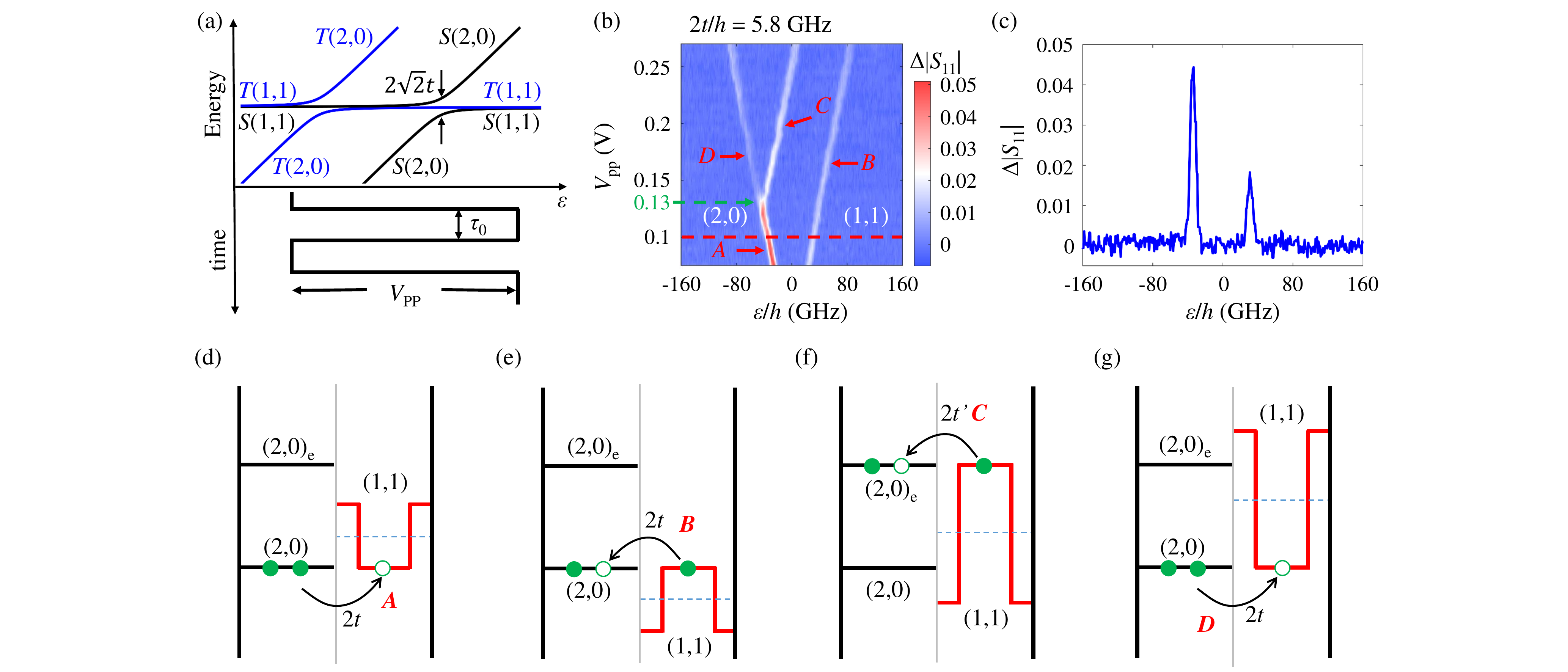}
\caption{
(a) Energy levels of two-electron spin states as a function of detuning $\varepsilon$.
A pulse train is applied to gate R which is generated from an arbitrary wavefunction generator with width $\tau_0$ and peak-to-peak voltage $V_{\text{pp}}$.
(b) Reflection signal $\Delta|S_{11}|$ measured as a function of detuning $\varepsilon$ and pulse voltage $V_{\text{pp}}$ after subtracting the background in the Coulomb blockade regime. 
(c) Horizontal line cut of the spectrum at $V_{\text{pp}} = \SI{0.1}{\V}$. 
(d)--(g) Schematic electrochemical potential diagrams for different transition lines in (b) labeled A, B, C and D.}
\label{fig:spectrum}
\end{figure*}

\sect{Experiment}
To perform the excited-state spectroscopy on the DQD, we apply a 50\% duty-cycle square pulse train to gate R with width $\tau_0$ and peak-to-peak voltage $V_{\text {pp}}$ [Fig.~\ref{fig:spectrum}(a)]. 
We sweep the detuning energy $\varepsilon$ through the (2,0)--(1,1) transition line by varying $V_\text{R}$, while the  applied pulse train modulates the detuning periodically.  
Then, the resonator response $\Delta|S_{11}|$ is measured as a function of $V_{\text{pp}}$ and $\varepsilon$ for $\tau_0=\SI{30}{\ns}$
(repetition rate $1/2\tau_0\approx\SI{16.67}{\MHz}$) and tunneling rate $2t/h\approx\SI{5.8}{\GHz}$ with probe frequency $f_{\text p}=f_{\text r}$.

Figure~\ref{fig:spectrum}(b) shows the spectrum after subtracting the background in the Coulomb blockade regime. 
In the spectrum, three transition lines are clearly observed, which can be divided into four parts for explanation convenience, labeled A, B, C and D. 
The relevant dynamic processes are schematically illustrated in 
Fig.~\ref{fig:spectrum}(d)--(g).
When the pulse voltage $V_\text{pp}$ is small, two slanting lines labeled A and B are observed. 
Line A (B) results from alignment of the electrochemical potentials of charge states (2,0) and (1,1) during the lower (upper) voltage level of the pulse train. 
In these situations, the energy detuning $\varepsilon$ is compensated by the pulse voltage $\pm V_{\text{pp}}/2$ in a half period, allowing coherent interdot electron tunneling with a rate of $2t$.  
With increasing $V_{\text{pp}}$, line A splits into lines C and D at $V_\text{pp}=\SI{0.13}{\V}$. 
The pulse voltage is large enough that it enables the transition between the excited state $(2,0)_{\text e}$ and (1,1) during the upper voltage level of the pulse train, schematically shown in Fig.~\ref{fig:spectrum}(f). 
We can directly determine the energy splitting $\Delta E=\SI{330}{\micro\eV}$ between the ground state and excited state of (2,0) from the energy difference between lines B and C.

More interestingly, the amplitude of line A is much larger than that of line B in the spectrum for a small pulse voltage $V_{\text{pp}}$. 
The different resonator responses can be explained as a result of Pauli spin blockade considering the spin degrees of freedom. 
Figure~\ref{fig:spectrum}(a) shows the schematic energy diagram of the two-electron spin states.
In the (2,0) configuration, the ground-state configuration for $\varepsilon<0$, two electrons favor the spin-singlet state, denoted S(2,0).
The triplet states T(2,0) are energetically inaccessible for a small pulse voltage ($V_{\text{pp}} < \SI{0.13}{\V}$).
For $\varepsilon>0$, the separated (1,1) charge configuration is dominant. All the spin states are accessible in the absence of an external magnetic field: the singlet, denoted S(1,1), and three triplets, denoted T$_{0,\pm}$(1,1).
These spin states could be strongly mixed in the (1,1) configuration due to the coupling of GaAs nuclear spins within the spin-dephasing time $3-\SI{10}{\ns}$ \cite{Johnson2005a,Petta2010a}.
Since the pulse frequency is much smaller than the dephasing rate, we suppose that all four spin states are fully mixed during the pulse period of large negative voltage.

For line A, the ground state is S(2,0) during the higher voltage level of the pulse train, schematically shown in
Fig.~\ref{fig:spectrum}(d). 
With a subsequent lower voltage level pulse, the transition between S(2,0) and S(1,1) occurs. 
In contrast, for line B, the ground charge state is (1,1) during the lower voltage level in 
Fig.~\ref{fig:spectrum}(e) and all the spin states S(1,1) and T$_{0,\pm}$(1,1) are accessible. 
Nevertheless, a subsequent higher voltage level pulse only allows the
S(1,1)--S(2,0) transition while the T(1,1)--S(2,0) transition is forbidden due to spin blockade \cite{Ono2002a,Johnson2005c}. 
Therefore, when performing the spectroscopy with a small voltage pulse train, only the singlet state contributes to the resonator response at the interdot transition, which strongly reduces the amplitude of line B in comparison to that of line A. 
Figure~\ref{fig:spectrum}(c) shows a horizontal cut of the energy spectrum through lines A and B at $V_{\text{pp}}=\SI{0.1}{\V}$. 
The height of the left peak corresponding to line A is much higher than that of the right peak corresponding to line B, which results from spin blockade.

\begin{figure}[ht]
\includegraphics[width=\columnwidth]{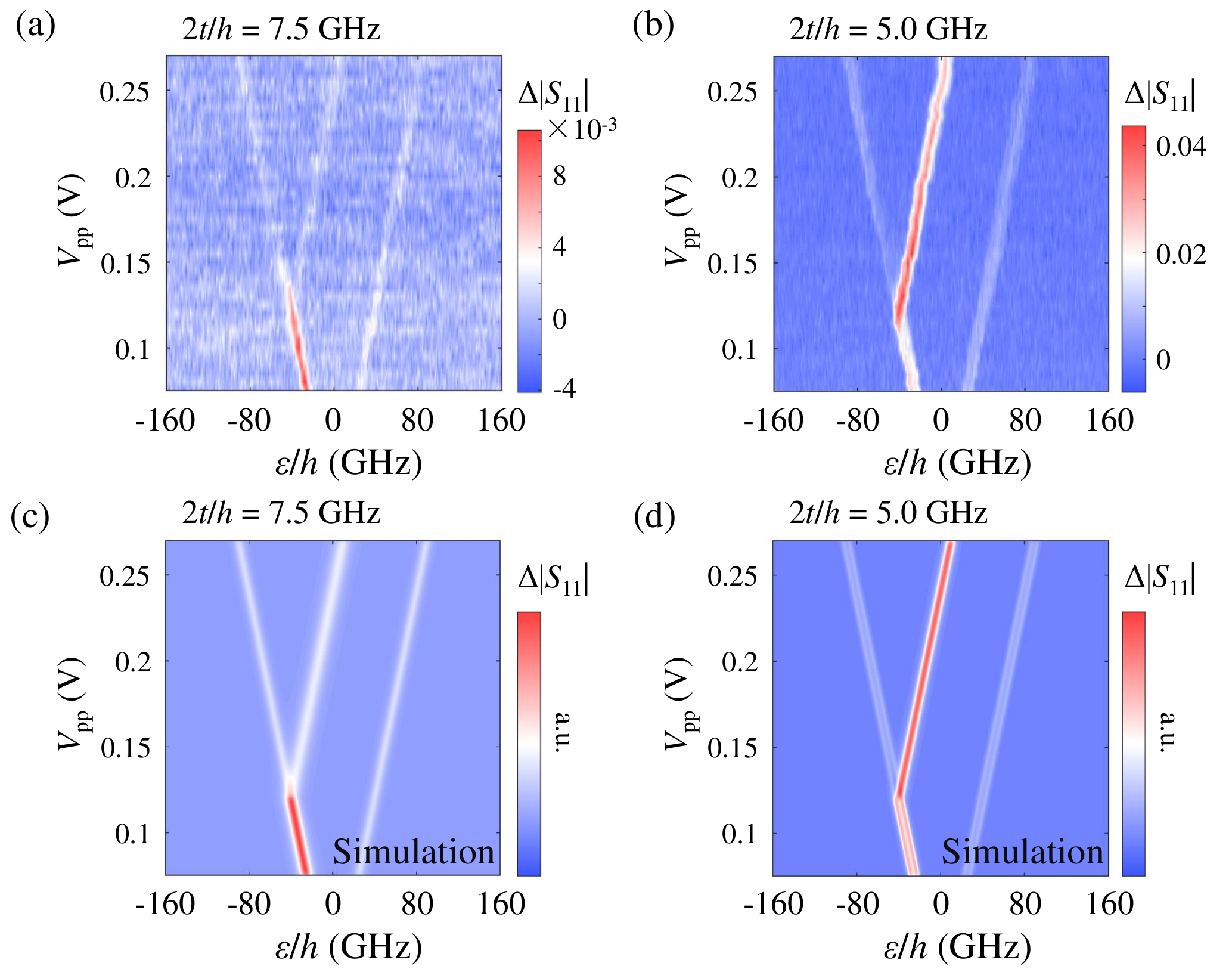} 
\caption{
(a)--(b) Excited-state spectroscopy for $2t/h = \SI{7.5}{\GHz}$ and $\SI{5.0}{\GHz}$, respectively. 
(c)--(d) Simulated results for the parameters used in panels (a) and (d).
}
\label{fig:3}
\end{figure}

When the pulse voltage $V_{\text{pp}}$ is larger than $\SI{0.13}{\V}$, the three triplet states T$_{0,\pm}$(2,0) are no longer inaccessible. 
The transition between T(2,0) and T(1,1) with a tunneling rate of $2t'$ is allowed when the two electrochemical potentials are in alignment by applying the pulse train [Fig.~\ref{fig:spectrum}(f)], labeled C in the excited-state spectrum. 
Line D, which corresponds to the S(2,0)--S(1,1) transition, is also influenced by the participation of the excited state T(2,0). Because of the large peak-to-peak pulse voltage $V_{\text{pp}}$, electrons can be loaded into S(2,0) or T(2,0) during the higher voltage level of the pulse train [Fig.~\ref{fig:spectrum}(g)].
Considering that only the electrochemical potentials of S(1,1) and S(2,0) are aligned during the lower voltage level, triplet states are not involved in the transition. 
Consequently, the resonator signal $|S_{11}|$ of line D is suppressed compared to that of line A.
Therefore, the dispersive photonic probe with pulsed-gate technology provides an effective readout method for singlet-triplet qubits.

Next, we investigate the $2t$ dependence of the excited-state spectrum by tuning the voltage bias on gate U.  
The results are shown in Fig.~\ref{fig:3}(a) and (b) for $2t/h = 7.5$ and $\SI{5.0}{\GHz}$, respectively, exhibiting different signal-to-noise ratios.
When $2t/h > f_{\text r}$, the qubit energy $\hbar\Omega=\sqrt{(\varepsilon^2+(2t)^2)}$ is always larger than the resonance frequency $hf_{\text r}$. The response amplitude to the interdot electron transition decreases rapidly with increasing $2t$ (see appendix for more details).
In Fig.~\ref{fig:3}(a), $2t/h = \SI{7.5}{\GHz}$ that is far beyond the resonator frequency $f_{\text r} = \SI{6.04}{\GHz}$, leading to a small signal-to-noise ratio.
In contrast, when $2t/h < f_{\text r}$, the DQD energy matches the resonator frequency at $\varepsilon=\pm\sqrt{(hf_{\text r})^2-(2t)^2}$ and the frequency shift $\Delta f_{\text r}$
reaches a maximum, which results in a good resonator response in Fig.~\ref{fig:3}(b).

Moreover, the resonator response to the (2,0)\textsubscript{e}--(1,1) transition in Fig.~\ref{fig:3}(b) is much stronger than that in the case of $2t/h = 5.8$ and $\SI{7.5}{\GHz}$. 
Considering that the excited state has a larger spatial distribution of the wavefunction and a larger overlap, the tunneling rate $2t'$ at the (2,0)\textsubscript{e}--(1,1) transition is larger than $2t$.  
For $2t/h = \SI{5.0}{\GHz}$, the corresponding $2t'$ is closer to the resonance frequency $f_{\text r}$.
As a result, the resonator is more sensitive to the (2,0)\textsubscript{e}--(1,1) tunneling process, with a strong enhancement of the amplitude of the middle transition line in Fig.~\ref{fig:3}(b). 
To examine our explanation, we theoretically simulate the experiment as shown in Fig.~\ref{fig:3}(c) and (d) based on our model (see appendix), and we achieve good agreement with the expeirment.

\sect{Discussion and summary}
The resonator-detected spectrum has a clean background because the resonator only responds to the elastic interdot tunneling that approaches the resonance frequency.
The resonator maintains a good response in the few-electron regime. 
To further improve the readout quality, suppressing microwave leakage to gate electrodes by using delicate filtering circuits \cite{Mi2017a,Harvey-Collard2020a} and improving the signal-to-noise ratio by the application of Josephson parametric amplifiers \cite{Stehlik2015a} are demanded. 

In summary, we have demonstrated the energy spectrum measurement of a four-electron GaAs DQD using a high-impedance superconducting NbTiN resonator by applying  50\% duty-cycle pulses to a gate electrode.
We attribute the different resonator responses to electron transitions in the spectrum to different combinations of aligned spin states and the effect of the Pauli exclusion principle. 
By tuning the voltage bias on the barrier gate, the influence of the interdot tunneling rate is also investigated and verified based on our model. 

Our results provide a useful means of sensitive readout for singlet-triplet qubits in cQED architectures.
This gate-based readout method is not only available for gate-defined DQD devices, but also can be easily extended to other cavity-coupled hybrid systems for spectroscopically probing qubits and characterizing energy levels. 

\begin{acknowledgments}
This work was supported by the National Key Research and Development Program of China (Grant No.\ 2016YFA0301700), the National Natural Science Foundation of China (Grants No.\ 61922074, 11674300, 61674132, 11625419 and 11804327), the Strategic Priority Research Program of the CAS (Grant No.\ XDB24030601), the Anhui initiative in Quantum Information Technologies (Grant No.\ AHY080000).
This work was partially carried out at the University of Science and Technology of China Center for Micro and Nanoscale Research and Fabrication.

\end{acknowledgments}

\clearpage
\appendix
\section*{Appendix}
\subsection{Theoretical method}
In absence of an external magnetic field, all three triplets T$_\pm$ and T$_0$ are degenerate. 
Then, for simplicity, the Hamiltonian for a two-electron double quantum dot system that conserves electron spins can be described in the singlet and triplet states basis 
[S(1,1), S(2,0), T(1,1) and T(2,1)] as
\begin{equation}
H_0=\left(
\begin{matrix}
0 & \quad t & \quad 0 & 0\\
t & \quad\varepsilon & \quad 0 & 0\\
0 & \quad 0 & \quad 0 & t'\\
0 & \quad 0 & \quad t' & \varepsilon +\Delta E
\end{matrix}
\right).
\end{equation}
Here, $t$ ($t'$) is the interdot coupling rate between singlet states, S(1,1) and S(2,0) [triplet states, T(1,1) and T(2,0)], and $\Delta E$ is the energy splitting between S(2,0) and T(2,0).

Applying a slow-frequency pulse train onto the DQD, the detuning energy is time-periodic,
$$
\varepsilon (\tau) \rightarrow \left\{
\begin{matrix}
\varepsilon + \alpha V_\text{pp}/2,& 2n\tau_0\leq \tau<(2n+1)\tau_0,\\
\varepsilon - \alpha V_\text{pp}/2,& (2n+1)\tau_0\leq \tau<(2n+2)\tau_0,
\end{matrix}
\right. 
$$
with an integer $n$ and a repetition rate of $1/2\tau_0$, where $2\tau_0$ is the pulse period, 
$V_\text{pp}$ is the peak-to-peak pulse voltage and $\alpha$ is voltage-to-energy conversion ratio of the pulse.
To simplify the contribution from the electron reservoir coupling to the DQD electron states, we assume that all the accessible two-electron spin states can be loaded with equal probabilities when the electrochemical potential of the dot lies below the electron reservoir. 
Hence for the (1,1) charge configuration, all the spin states can be initialized with a probability of one fourth. 
For the (2,0) charge configuration, however, the initialization depends on the pulse amplitude. 
For a small driving amplitude $V_\text{pp}$, only S(2,0) is accessible. 
If $V_\text{pp}$ is large enough, which in our experiment is $\SI{0.13}{\V}$, triplet states T$_{0,\pm}$(2,0) can also be accessed.
Since the pulse frequency is much slower than the decoherence rate \cite{Johnson2005a}, the cavity reflection signal can be treated as an average response to the stationary state occupation for the two different pulse levels. 
Given the input-output  theory \cite{Viennot2015a,Mi2018a,Landig2019a}, 
we then simulate the exited-state spectra for the parameters used in the experiment as shown in Fig.~\ref{fig:3}(c) and (d). 
The relevant parameters are the coupling strength $g_0/2\pi\approx\SI{104.5}{\MHz}$, photon internal, external, total decay rate $(\kappa_i,\kappa_e,\kappa)/2\pi=(4.8,30.5,35.3)\SI{}{\MHz}$, qubit decoherence rate $\gamma/2\pi\approx\SI{72}{\MHz}$, and interdot tunneling rates $2t/h=\SI{7.5}{\GHz}$ and $\SI{5}{\GHz}$.
All the simulation plots are convoluted with a Gaussian of width $\SI{4}{\micro\eV}$, considering an inhomogeneous broadening of the excited-state spectrum induced by the slow uctuations of the detuning 
\cite{Forster2014a,Cao2013b}.
Because of the difficulty in extracting $t'$ and the triplets-resonator coupling strength $g_0'$, we assume that  $2t'=\SI{6}{\GHz}$ for $2t=\SI{5}{\GHz}$ and  $2t'=\SI{10}{\GHz}$ for $2t=\SI{7.5}{\GHz}$ due to a larger spatial wavefunction of the excited states, and $g_0'\approx g_0$.
The results in Fig.~\ref{fig:3}(c) and (d) agree with the experiment qualitatively well, which reveal that our measurement can also be used to estimate the tunneling rate between excited states.

\subsection{Influence of other system parameters}
According to Hamiltonian \eqref{dispersive}, the resonator frequency is related to the coupling strength $g_{\text c}$ and resonator-DQD energy difference $\Delta$, which are both functions of the interdot tunneling rate $2t$. 
To explain the different signal-to-noise ratios in Fig.~\ref{fig:3}, we calculate the frequency shift dependence on $2t$ and $\varepsilon$ by solving the Hamiltonian. 

\begin{figure}[bth]
\includegraphics[width=\columnwidth]{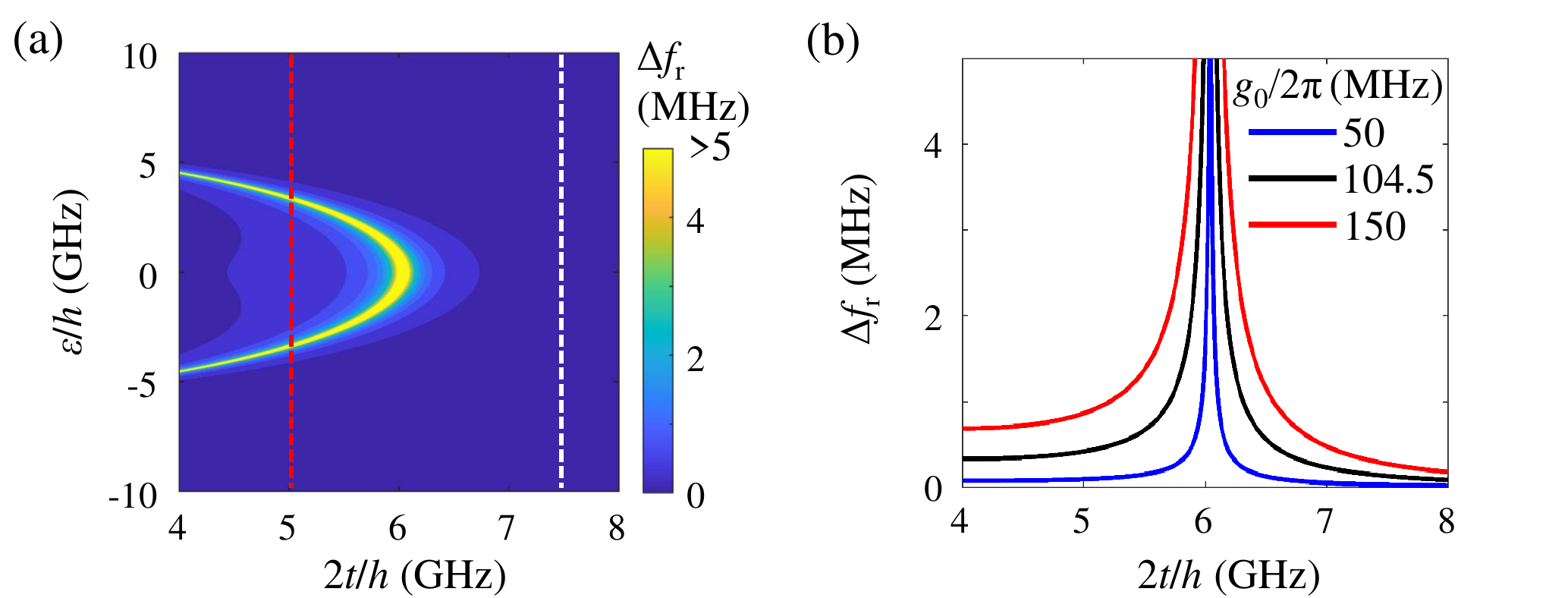} 
\caption{
(a) Frequency shift $\Delta f_{\text r}$ as a function of $2t$ and $\varepsilon$. 
The white and red dashed lines correspond to  the values of $2t$ in Fig.~\ref{fig:3}(a) and (b), respectively.
Other parameters are the same as those extracted from the experiment.
(b) Frequency shift $\Delta f$ as a function of $2t$ for coupling strengths
$g_0/2\pi=\SI{50}{\MHz}$ (blue line), $g_0/2\pi=\SI{104.5}{\MHz}$ (black line) and $g_0/2\pi=\SI{150}{\MHz}$ (red line), along the zero detuning in panel (a).
}
\label{fig:4}
\end{figure}

As shown in Fig.~\ref{fig:4}(a), the two dashed lines colored in red and white correspond to the values of $2t$ in Fig.~\ref{fig:3}(a) and (b), respectively.
When $2t/h > f_{\text r}$, the qubit energy $\Omega=\sqrt{(\varepsilon^2+(2t)^2)}$ is always larger than the resonance frequency $hf_{\text r}$ and the frequency shift $\Delta f_{\text r}$ decreases rapidly with increasing $2t$. 
Consequently, for $2t/h = \SI{7.5}{\GHz}$, the large tunneling rate which is far beyond the resonator frequency $f_\text{r}=\SI{6.04}{\GHz}$, 
leading to a small signal-to-noise ratio in Fig.~\ref{fig:3}(a).
When $2t/h < f_{\text r}$, the DQD energy matches the resonator frequency at $\varepsilon=\pm\sqrt{(hf_{\text r})^2-(2t)^2}$.
Near both reonsnace points, $\Delta f_{\text r}$
reaches a maximum, which results in a good resonator response in Fig.~\ref{fig:3}(b).
Therefore, the excited-state spectrum for $2t=\SI{5}{\GHz}$ is much better than that for $2t=\SI{7.5}{\GHz}$.
Note that the distance between the two resonance points is much  smaller than other energy scales in the spectrum, so the splitting of peaks is not clearly visible in Fig.~\ref{fig:3}(b) but results in bolder transition lines than those in Fig.~\ref{fig:3}(a).

In addition to the tunneling rate, the charge-photon coupling rate $g_0$ 
also has an influence on the resonator response. 
Based on the numerical results in Fig.~\ref{fig:4}(b), we determine the $2t$ dependence of $\Delta f_{\text r}$ at zero detuning for different coupling strengths $g_0$. 
Obviously, a larger $g_0$ leads to a larger $\Delta f_{\text r}$ as well as a larger amplitude $|S_{11}|$.

\clearpage

\bibliography{ref}

\end{document}